\documentclass{PoS}
\usepackage{graphicx}
\usepackage{epsf}
\usepackage{color}
\usepackage{natbib}
\usepackage{fancyhdr}
\newcommand{\getlength}[1]{\ifx#1\end \let\next=\relax
            \else\advance\count255 by1 \let\next=\getlength\fi \next}
\newcommand{\ifnularg}[1]{ \count255=0 \getlength#1\end \ifnum\count255=0 }
\newcommand{\ifm}{\makebox{}\ifmmode}
\long\def\ifundefined#1#2#3{\expandafter\ifx\csname
  #1\endcsname\relax#2\else#3\fi}
\newcount\switch
\newcommand{\Endmat}{\ifnum\switch=0$\fi}
\newcommand{\beq}   { \begin{eqnarray} }
\newcommand{\eeq}[1]{ \ifnularg{#1} \end{eanarray} \else
                      \label{#1}\end{eqnarray}    \fi }
\newcommand{\eeql}   { \end{eqnarray} }
\newcommand{\eeqn}   { \nonumber \end{eqnarray} }

\newcommand{\dss}{\displaystyle}

\newcommand{\Frac}[2]{\frac{\displaystyle\strut #1}{\displaystyle\strut #2} }

\newcommand{\upp}[1]{ {}^#1{\scriptstyle\kern-0.3em . \kern0.15em } }

\newcommand{\lp}{ \left(  }
\newcommand{\rp}{ \right) }

\ifundefined{sun}{}{}

\renewcommand{\epsilon}{\varepsilon}
\renewcommand{\phi}{\varphi}

\renewcommand{\refname}{\par\vspace{-2ex}\par}
\definecolor{Dblue}{rgb} {0.152,0.268,0.720}
\definecolor{Dred}{rgb}{0.535,0.148,0.070}
\definecolor{Dgreen}{rgb}{0.094,0.410,0.094}
\newcommand{\Blb}[1]{\textcolor{Dblue}{\bf #1}}
\newcommand{\Grb}[1]{\textcolor{Dgreen}{\bf #1}}
\newcommand{\Rdb}[1]{\textcolor{Dred}{\bf #1}}
\definecolor{LightSiren}{rgb}{0.900,0.850,0.980}
\definecolor{LightGreen}{rgb}{0.900,0.980,0.850}

\newcommand{\ee}{\everymath=\expandafter{\the\everymath\scriptstyle}}

\title{Modeling of path delay in the neutral atmosphere: \enskip a paradigm shift}
\ShortTitle{Modeling of path delay in the neutral atmosphere}
\FullConference{12th European VLBI Network Symposium and Users Meeting,\\
                 7-10 October 2014\\
                 Cagliari, Italy}

\author{\speaker{Leonid Petrov}\\
       {Astrogeo Center, Falls Church, VA, USA}\\
        E-mail: \email{Leonid.Petrov@lpetrov.net}}

\abstract{
Computation of propagation effects in the neutral atmosphere, namely
path delay, extinction, and bending angle is a trivial task provided the 
4D state of the atmosphere is known. Unfortunately, the mixing ratio of water
vapor is highly variable and it cannot be deduced from surface measurements.
That fact led to a paradigm that considers path delay and extinction in
the atmosphere as a~priori unknown quantities that have to be evaluated 
from the radio astronomy data themselves. Development of our ability to
model the atmosphere and to digest humongous outputs of these models that
took place over the course of the 21st century changed the game. Using the 
publicly available output of operational numerical weather model GEOS
run by NASA, we are in a position to compute path delay through the neutral 
atmosphere for any station and for any epoch from 1979 through now with 
accuracy of 45 ps $*$ cosec elevation. We are in a position to compute 
extinction with accuracy better than 10 pro cents. We are in a position 
to do it routinely, in a similar way how we update apparent star positions 
for  precession and nutation. Moreover, we are in a position to do it now.
As a demonstration of current capabilities, I have computed time series of
path delays for aall radiotelecopes that I was aware of (220 sites) since 1979 
with a step 3-6 hours. Results of the validation tests are presented. A new 
paradigm of data analysis assumes that we know the atmosphere propagation 
effects a priori with the accuracy higher that one could deduce them from 
radio astronomy observations. 
}

\begin{document}

\section{Introduction}
\par\vspace{-1ex}

  Radio waves travel billions years from typical sources observed with VLBI.
At the very end of their journey, at the last millisecond, they propagate 
through the Earth's atmosphere. While multi-frequency observations allow us
to evaluate the contribution of the ionized component of the Earth's 
atmosphere, modeling propagation effects in the neutral atmosphere poses 
a challenge. There are four effects: 1)~refraction, i.e. trajectory bending, 
2)~delay in the atmosphere, 3)~attenuation, and 4)~atmosphere emission. 
If we know the state of the atmosphere, we can compute these quantities. 
The crux of the problem is that the state of the 4D atmosphere cannot be deduced 
from surface measurements. The reason of that is water is underwent phase 
transition at the atmospheric layers that are above the surface, and 
as a result, the mixing ratio of water vapor is highly volatile.

\section{Old paradigm}
\par\vspace{-1.0ex}

  The old paradigm assumed we do not know the instantaneous state of the 
atmosphere. In order to circumvent the lack of knowledge, some greatly
simplified or just wrong a~priori models were used for computation of path 
delay in the past. In particular, \citet{r:Saa72} model became very popular. 
That model, in a way how it was implemented, assumed that water vapor present 
in the atmosphere had thermodynamic properties of dry air. A typical error 
of path delay computed that way was 300--1300~ps, i.e. 5--15\% of the total 
delay. In order to improve modeling, the astronomical data themselves were 
suggested to be used for solving residual parameters of the propagation model. 
It is assumed in the framework of this paradigm that the path delay 
$\tau(t,e,A)$ {\it can be}\footnote{Unfortunately, it is often forgotten, that 
such a decomposition is a only a simplified approximation within its range of
applicability.} decomposed into two azimuth-independent
components:
\beq
  \tau(t,e,A) = \tau_d(t) \cdot m_d(e) \enskip + \enskip t_w(t) \cdot m_w(e) ,
\eeq{e:e1}
  where $\tau_d(t)$ is the so-called dry component, $m_d(e)$ is the so-called dry
mapping function, i.e. a derivative with respect to elevation; $t_w(t)$ and
$m_w(e)$ is wet path delay and its derivative with respect to elevation angle;
$t$, $e$, and $A$ stands for time, elevation angle above the horizon,
and azimuth.

  For evaluation of extinction in the atmosphere measurements of antenna
brightness temperature at different elevations were often made (tipping 
curves).

\section{New paradigm}

  Advances in numerical weather models made it possible to evaluate parameters
of the atmosphere using various ground, air-born, and space-born measurements
that are assimilated into a dynamic model of the atmosphere. The output of 
these models defines the parameters of the state of the atmosphere on 
a 4-dimensional grid. Three parameters are important for reduction of astronomy 
observations: air temperature $T$, total atmospheric pressure $P$, and 
partial pressure of water vapor $P_w$. Thus, the state of the atmosphere in
the new paradigm is considered known. At the moment, there are several centers
in the world that produce numerical models of the atmosphere. I used models 
produced by the NASA Global Modeling and Assimilation Office 
(GMAO)\footnote{The output of the models is available at 
http://gmao.gsfc.nasa.gov} for this study.

\begin{table}
   \caption{GMAO models of the atmosphere}
   \par\medskip\par
   \begin{tabular}{llll}
     MERRA:    & Since 1979.01.01  & 
                 72 lev $\times 0.5^\circ \times 0.67^\circ \times 6^h$ & Latency: $40^d$ \\
     GEOS FPIT & Since 2000.01.01 & 
                 72 lev $\times 0.5^\circ \times 0.67^\circ \times 3^h$ & Latency: $12^h$ \\
     GEOS FP   & Since 2011.09.01 & 
                 72 lev $\times 0.25^\circ \times 0.3125^\circ \times 3^h$ & Latency: $12^h$ \\
   \end{tabular}
\end{table}

\section{Computation of path delay}

  Electromagnetic waves in the medium propagate with group velocity $v_m$ that
is always smaller than than the speed of light in vacuum $c$. Propagating 
through the medium that has variable refractivity defined as 
$r=\frac{c-v_m}{v_m}$, the electromagnetic wave changes its direction. 
Refractivity is a simple function of $P$, $T$, and $P_w$ that depends on 
coefficients measured in laboratory. According to \cite{r:apa11}, refractivity 
in radio range can be computed with accuracy $\sim\!\! 0.1\%$.

  There are several ways to solve for ray trajectory in a continuous medium 
with known refractivity field. I prefer an approach based on solving 
the variational problem: the actual trajectory that the electromagnetic wave 
travels is the one that minimizes travel time (Fermat principle). This principle
was first formulated empirically in 1662 but nowadays can be derived 
directly from Maxwell equations \citep[see, for instance][]{r:landau}. 
A general solution of this problem of calculus of variations was found by 
Euler in 1744 and it is reduced to solving a non-linear system of differential
equations of the 4th order with mixed initial values. The coefficients
of equations depend on the refractivity and its gradient.

  Considering propagation of electromagnetic wave through a continuous medium 
requires representation of refractivity $r(h,\lambda,\phi_{gd},t)$ as 
a continuous differentiable function. The mathematical model that I used 
for such a representation is an expansion of refractivity into a 4D tensor 
product of B-spline functions of the $m$th degree $B^m_p$:
\beq
     r(h,\lambda,\phi_{gd},t) = \sum_{i=1-m}^{i=d_1-1}
                                \sum_{j=1-m}^{j=d_2-1}
                                \sum_{k=1-m}^{k=d_3-1}
                                \sum_{l=1-m}^{l=d_4-1}
                                 f_{ijkl} \,
                                 B^{m}_i(h) \, 
                                 B^{m}_j(\lambda) \, 
                                 B^{m}_k(\phi_{gd}) \, 
                                 B^{m}_l(t) .
\eeq{e:e2}

  The property of B-spline, minimal support, reduces the problem
of finding the coefficients $f_{ijkl}$ to solving a system of algebraic 
equations with an $m$-diagonal matrix, which can be solved extremely
efficiently at modern computers.

  Representation of the refractivity fields in the form of the
tensor product of B-splines of the 3rd degree (m=3), reduces 
differential equations that are a solution of the variational problem 
to a system of tri-diagonal algebraic non-linear equations. That system
is efficiently solved by iterations starting from the 1st approximation 
that radio waves propagate along the straight line. Two iterations are 
sufficient to reduce errors of a numerical solution below 1~ps.

  After determining the trajectory of the radio wave from the 
emitter to the receiver, we can easily find the time delay 
in the neutral atmosphere along the trajectory $\eta(\xi), \zeta(\xi)$ 
by integrating refractivity along the path:
\beq
     \tau_{na} = \strut 
                 \Frac{1}{c} \int_0^\infty \lp\lp 1 + r \lp \xi, \eta, \zeta \rp\rp
                 \sqrt{1 + \lp \Frac{d\eta}{d\xi}  \rp^2 +
                           \lp \Frac{d\zeta}{d\xi} \rp^2}  - 1 \rp d\xi ,
\eeq{e:e3}
  where the Cartesian coordinate system $\xi,\eta,\zeta$ is chosen in such 
a way that axis $\xi$ is along the straight line from the receiver to the 
emitter, and axes $\eta, \zeta$ are orthogonal to that direction.

  Attenuation in the atmosphere can be computed by evaluating the integral
that resembles \ref{e:e3}:
\beq
  a(\xi_0) = \int_{\xi_0}^\infty \alpha(P,P_w,T,f) 
                 \sqrt{1 + \lp \Frac{d\eta}{d\xi}  \rp^2 +
                           \lp \Frac{d\zeta}{d\xi} \rp^2}  d\xi ,
\eeq{e:e4}
  where $\alpha(P,P_w,T,f)$ is the specific attenuation coefficient that has
a strong dependency on frequency $f$. In particular, attenuation of the signal 
at the ground is $a(0)$.

  Finally, atmosphere brightness temperature $T_{atm}$ is found by solving
the radiative transfer equation, which is reduced to evaluation of the integral
\beq
  T_{atm}=  \dss \int_0^\infty T \lp \xi, \eta, \zeta \rp
            \cdot e^{-a(\xi)}
            \sqrt{1 + \lp \Frac{d\eta}{d\xi}  \rp^2 +
                      \lp \Frac{d\zeta}{d\xi} \rp^2} d\xi .
\eeq{e:e5}

\begin{figure}[ht]
   \caption{ Baseline length repeatability for a case when residual 
             zenith path delay is solved for (\Rdb{Left}) and when 
             no path delay was estimated (\Grb{Right}). The baseline 
             length repeatability extrapolated to the Earth's diameter 
             is 1.12~cm for the first case and 2.32~cm for the second 
             case.
           }
   \begin{center}
      \includegraphics[width=0.47\textwidth]{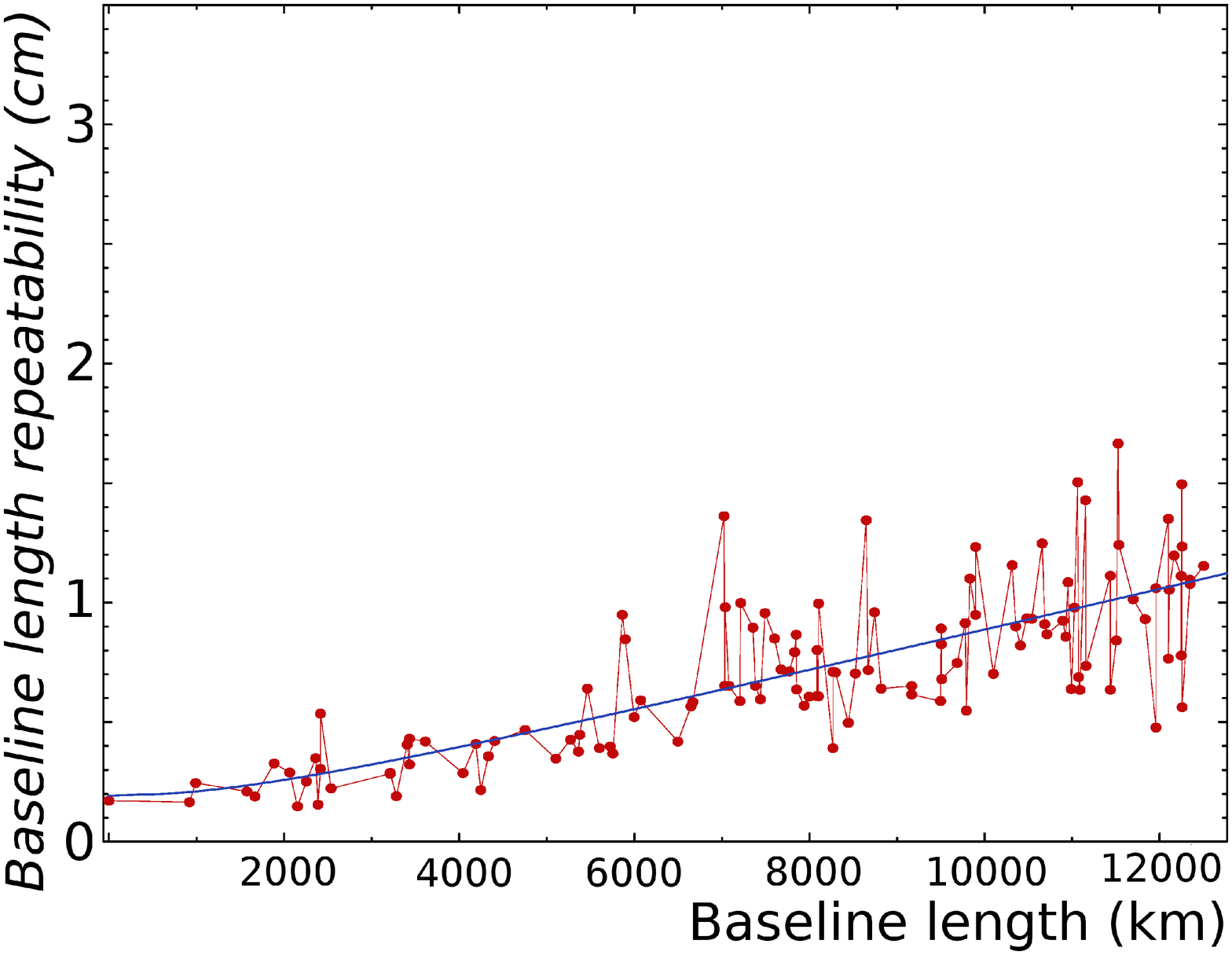}
      \hspace{0.03\textwidth}
      \includegraphics[width=0.47\textwidth]{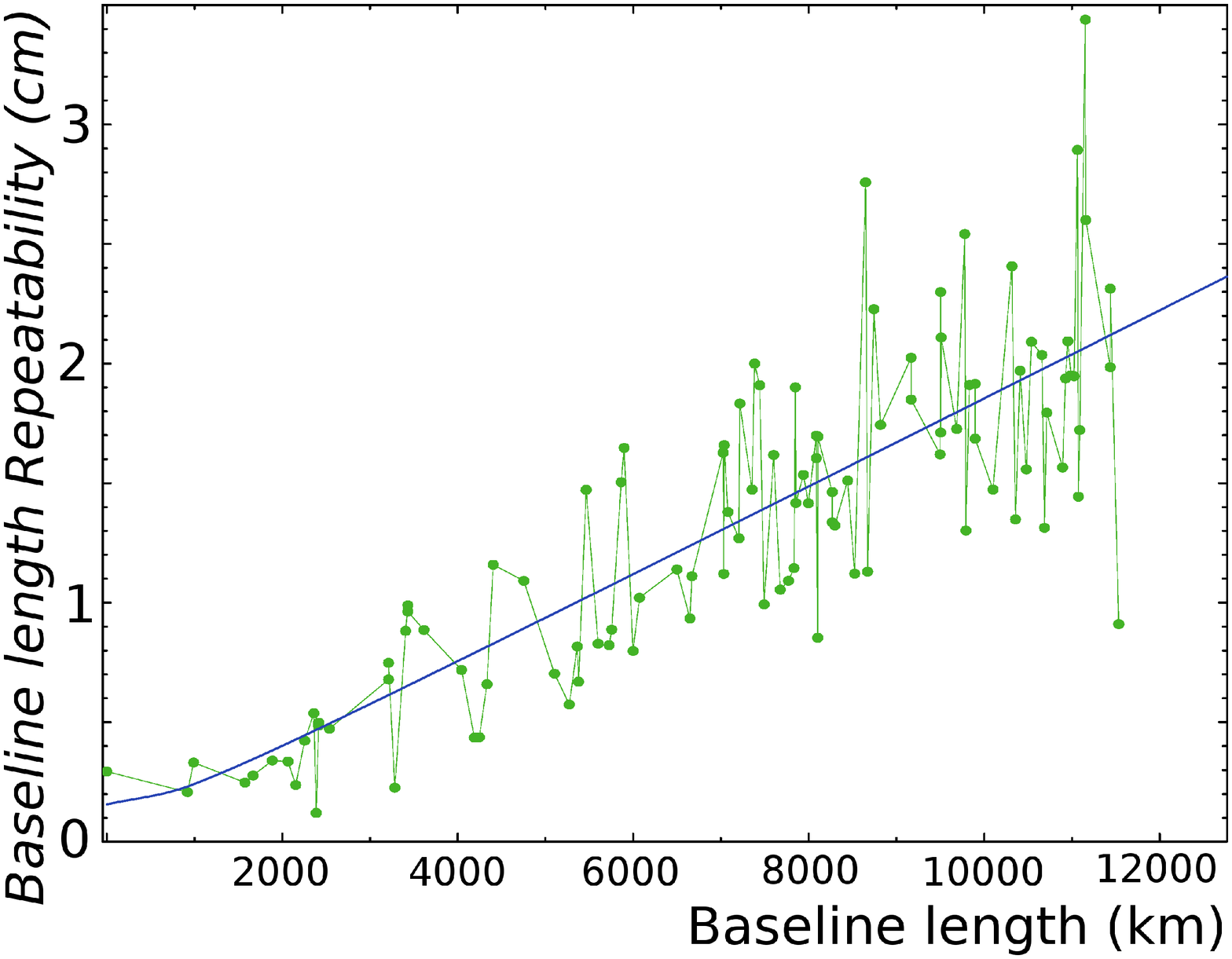}
   \end{center}
   \label{f:basrep}
\end{figure}

  At a given station, given epoch, integrals $\tau_{na}$, $a$, and $T_{atm}$ 
are functions of azimuth and elevation. For logistical reasons it is 
convenient to compute these quantities on a 2D azimuth-elevation regular 
grid (that does not have to be uniform), expand them into the B-spline basis
and store the coefficients. This approach disentangles computation of
integrals \ref{e:e3}--\ref{e:e5} from radio astronomy data analysis.

\section{Results}

  I computed azimuth-elevation coefficients for path delay expansions for all 
220 VLBI stations that participated in radio astronomy observations since 
1979 through present using GMAO numerical weather models GEOS-FPIT and MERRA. 
The input dataset has rather significant size, 36Tb, but still manageable.

  In order to evaluate the accuracy of path delays computation, I ran two 
solutions: with estimating residual path delay in zenith direction and 
without estimating. For each baseline I estimated the weighted root mean 
square (wrms) of deviations of baseline lengths with respect to the linear 
model --- the so-called baseline length repeatability test. The baseline 
length repeatability dependence on baseline length $L$ (blue lines in 
Figure~\ref{f:basrep}) was fitted to $R(L) = \sqrt{A^2 + (B \cdot L)^2}$, 
where $A$ and $B$ are fitted coefficients. When the baseline length between 
two stations is approaching to the Earth's diameter, its vector is 
approaching to the local vertical at both stations. Therefore, the wrms 
of station position uncertainties can be evaluated as 
$R(L=\mbox{diam})/\sqrt{2}$. Comparing the baseline length repeatabilities 
extrapolated to the Earth's diameter from two solutions, I derived the 
additional variance in vertical station positions due to errors in 
a~priroi path delays derived from the output of numerical weather 
models: 1.45~cm (48~ps).

  Another approach to evaluate errors of a~priori path delay is to compute
the rms of estimates of residual path delay in zenith direction. It varies
from 0.6~cm at polar station {\sc nyales20} to 2.2~cm at tropical station 
{\sc sc-vlba} with an average value of 1.25~cm (37~ps). It is instructive 
to note that the average zenith total path delay is 251.4~cm and the average 
zenith wet path delay is 12.3~cm. This allows me to conclude that total path 
delay can be computed with accuracy 0.5\%, and the contribution of water vapor
to path delay can be computed with accuracy 10\%.

\begin{figure}[ht]
   \caption{Atmosphere brightness temperature at elevation $45^\circ$ 
            at SARDINIA at 86.304 GHz}
   \begin{center}
      \includegraphics[width=0.66\textwidth]{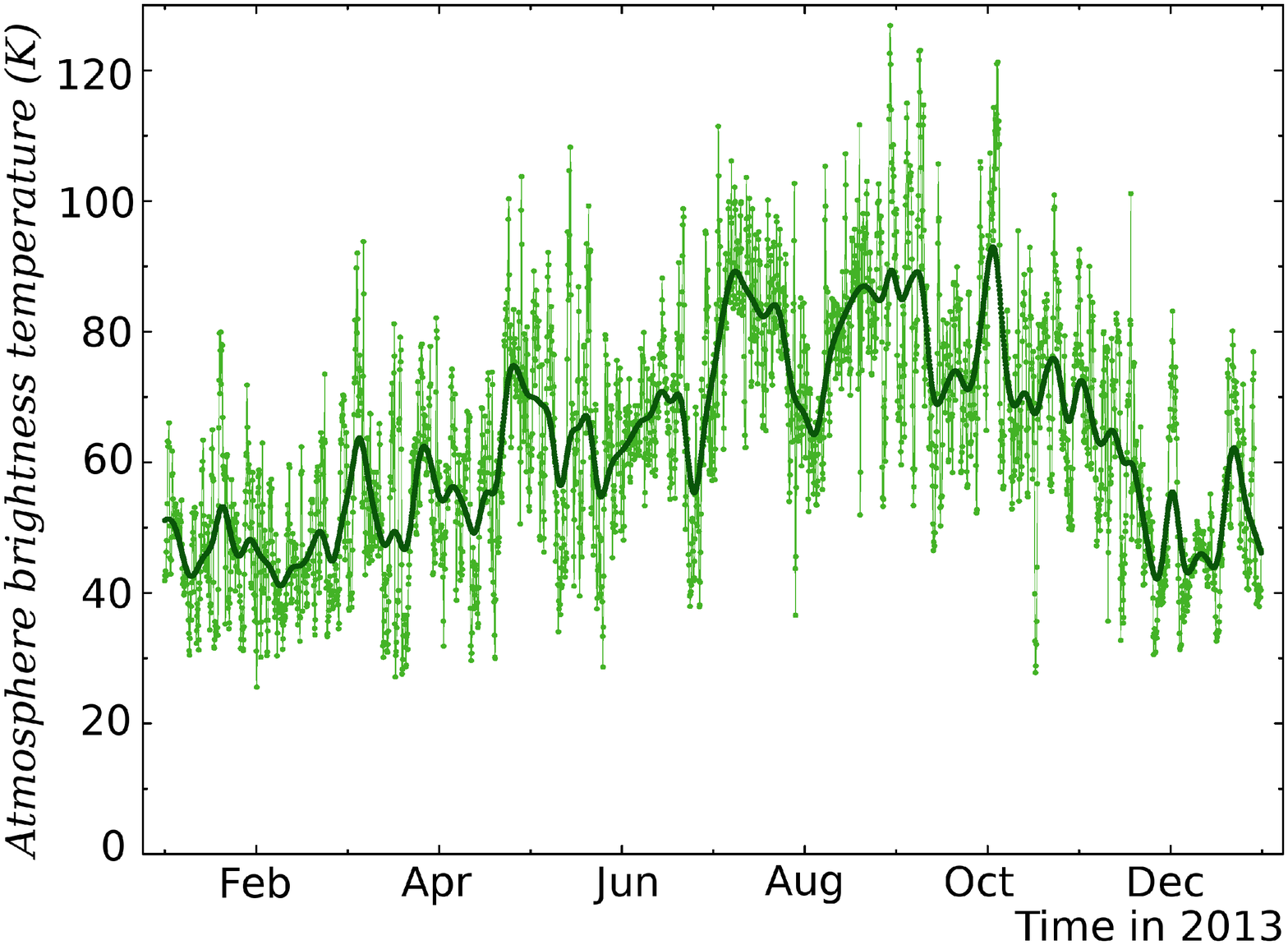} 
   \end{center}
   \label{f:srt}
\end{figure}

  Figure~\ref{f:srt} shows the estimates of the atmosphere brightness 
temperature at SARDINIA at 86.304 GHz. This figure may help us to make a
rough prediction of the atmosphere contribution to station performance. 
We see that a naive surmise that the performance in winter will be always 
significantly better than in summer is a gross simplification.

  Using estimates of the atmosphere attenuation from the output of numerical 
weather models is straightforward: we just calibrate fringe amplitudes 
by dividing them by $e^{-a}$ factor and thus relate them to the top of 
the atmosphere.

\begin{figure}[ht]
   \caption{ K-band system temperature in K as a function of time in sec
             (\Blb{Left}) and receiver temperature (\Grb{Right}) after 
             subtraction of atmosphere brightness temperature computed 
             using the output of numerical weather model GEOS-FPIT.
           }
   \begin{center}
      \includegraphics[width=0.48\textwidth]{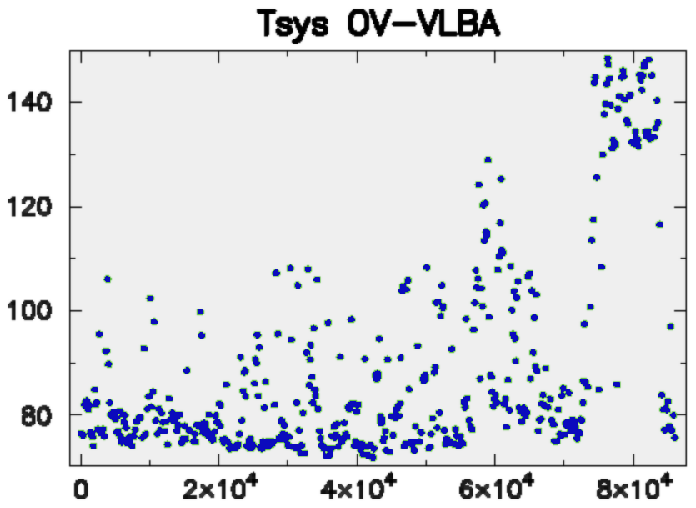}
      \hspace{0.03\textwidth}
      \includegraphics[width=0.47\textwidth]{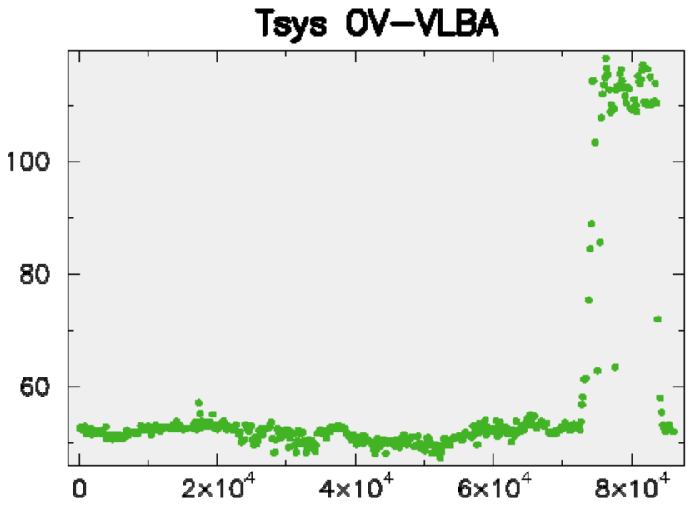}
   \end{center}
   \label{f:trec}
\end{figure}

  There are also certain benefits of being able to compute atmosphere 
brightness temperature using the output of numerical weather model. 
The difference between system temperature $T_{sys}$ and $T_{atm}$ is the sum 
of receiver temperature $T_{rec}$ and the spill-over term that does not 
depend on time, but depends on elevation and, in a less extent, on azimuth. 
If the receiver works properly, $T_{rec}$ is stable. Stacking 
$T_{sys}$ and $T_{atm}$ over the time range when $T_{rec}$ is stable,
we can separate the spillover term from $T_{rec}$ and develop an empirical
model for it. Analysis of $T_{rec}$ time series allows us to identify the 
period of time when it had anomalies (see Figure~\ref{f:trec}). This allows 
us to clean the data for bad estimates of $T_{sys}$ caused by radio 
interference and restore missing $T_{sys}$ by extrapolating $T_{rec}$.

  It is difficult to estimate errors of atmosphere brightness temperature
and attenuation in the atmosphere directly. It is reasonable to expect that 
in the frequency range where the dominating contribution to the atmosphere 
attenuation is water vapor, their errors are close to the uncertainty of 
wet path delay, i.e. around 10\%. Estimates of accuracies demonstrate
that certain old-fashion steps of data calibration, such as inserting 
geodetic blocks into schedules and measuring tipping curves, are nowadays
unnecessary, since even better accuracies can be achieved by utilizing 
the output of numerical weather models.

\end{document}